# Induced Electromotive Force (EMF) Profile in a Circular Loop Passing a Limited Rectangular Area with Perpendicular Uniform Constant Magnetic Field

Sparisoma Viridi[1], Nurul Khotimah[1], and Khairurrijal[2]

[1]Nuclear Physics and Biophysics Research Division
[2]Physics of Electronic Materials Research Division
Intitut Teknologi Bandung, Bandung 40132, Indonesia
dudung@fi.itb.ac.id, nurul@fi.itb.ac.id, krijal@fi.itb.ac.id

*Abstract*

Profile of induced eletromotive force (EMF) for a circular loop (CL) entering and leaving a limited rectangular area which has perpendicular uniform magnetic field is reported in this work. The influence of parameters of the sytem to the induced EMF profile is discussed.

### Introduction

The use of Lenz's law in determining direction of induction current in a loop entering our leaving a perpendicular magnetic field is already common in detail for rectangular loop in uniform magnetic field [1] or only as iillustration for circular loop in ununiform magnetic field along a long straight wire [2]. Detail of induced electromotive force (EMF) profile as function of time is preported in this work as a circular loop passing through a rectangular area with perpendicular uniform magnetic field, which can be used to improve previous reported work in measuring angular velocity of rotating disk [3].

### Induced EMF and current formulation

According to Faraday's law induced EMF $\varepsilon$ produced along a closed path can be formulated as [4]

$$\varepsilon = -\frac{d\Phi}{dt}, \quad (1)$$

with magnetic flux $\Phi$ is

$$\Phi = \oint \vec{B} \cdot d\vec{A}. \quad (2)$$

Perpendicular to element of area $d\vec{A}$ and uniform magnetic field $\vec{B}$ is chosen in this case, that will lead Equatio (2) into

$$\Phi = \oint B dA = B \oint dA = BA. \quad (3)$$

Then, Equation (1) will be simplified into

$$\varepsilon = -B\frac{dA}{dt}. \quad (4)$$

Induction current $I$ in a loop with area $A$ and resistance $R$ can be found from Ohm's law and Equation (4), which is

$$I = \frac{\varepsilon}{R} = -\frac{B}{R}\frac{dA}{dt}. \quad (5)$$

The minus sign in Equation (4) is from Lenz's law, where induction current produces magnetic field that opposes the change of magnetic flux that intially produces the induced EMF.

If the loop is constructed from a wire with diameter $D_w$, length $l$, and resistivity $\rho$, then with the resistance of the loop is [5]

$$R = \frac{4\rho l}{\pi D_w^2}. \quad (6)$$

Subtitue Equation (6) into Equation (5) will lead to

$$I = -\left(\frac{\pi D_w^2 B}{4\rho l}\right)\frac{dA}{dt}. \quad (7)$$

### Change of loop area

Equation (4) and (7) tell us that induced EMF and current can be obtained if we know how to calculate the change of loop area capturing the magnetic field.

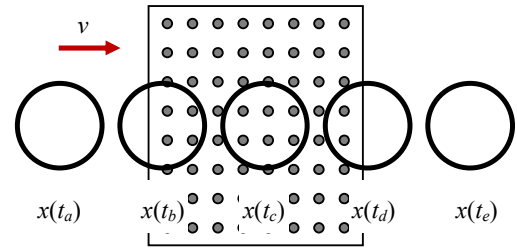

Figure 1. Position of circular loop at different time: $t_a < t_b < t_c < t_d < t_e$.

A circular loop (CL) with diameter $D_l$ is considered in this work, where its position for several times is illustrated in Figure 1. Position differences in the figure are related through

$$x(t_f) - x(t_i) = (t_f - t_i)v \quad (8)$$

for any initial $i$ and final $f$ position with $t_i < t_f$.

Change of area $A$ can be divided into three categories based on position of CL to the magnetic field area (MFA): (i) CL is fully outside and inside of MFA, (ii) CL is entering MFA, and (iii) CL is leaving MFA. In the first category there will be no change of area $A$, in the second the change will be positive, and in the last the change will be negative.

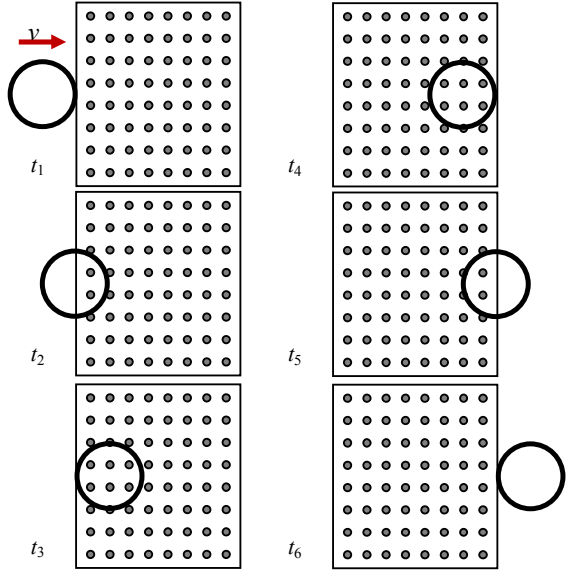

Figure 2. Position of CL as: entering MFA ($t_1$, $t_2$, $t_3$) and leaving MFA ($t_4$, $t_5$, $t_6$).

The first category is represented in three time intervals, which are

$$\frac{dA}{dt} = \begin{cases} 0, & t < t_1 \\ 0, & t_3 < t < t_4 \\ 0, & t > t_6 \end{cases}. \quad (9)$$

The formulation of change of area $A$ is different but similar for difference time interval. Illustration for second category in time interval $t_1 < t < t_2$ is given in Figure 3.

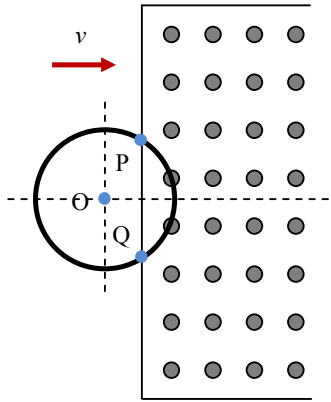

Figure 3. Position of CL as entering MFA for $t_1 < t < t_2$: with two points P and Q on the circumference.

For a time $t$ the CL can be rewritten from Figure 3 to give better picture how to obtain the area $A$ as shown in Figure 4. If the position of center of CL or point O is defined as $x$ and left side of MFA is defined as $x_L$ then it can be obtained from Figure 4 that

$$R = \frac{D_l}{2}, \quad (10)$$

$$A + A_{OPQ} = \frac{\theta}{2} R^2, \quad (11)$$

$$(x_L - x) = R \cos\left(\frac{\theta}{2}\right), \quad (12)$$

$$A_{OPQ} = R^2 \sin\left(\frac{\theta}{2}\right)\cos\left(\frac{\theta}{2}\right). \quad (13)$$

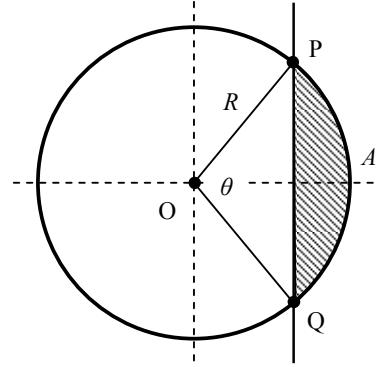

Figure 4. Area $A$ (shaded region) of CL as entering MFA for $t_1 < t < t_2$.

Then, it can be found that

$$\cos\left(\frac{\theta}{2}\right) = \frac{x_L - x}{R}, \quad (14)$$

$$\sin\left(\frac{\theta}{2}\right) = \frac{\sqrt{R^2 - (x_L - x)^2}}{R}. \quad (15)$$

From Equation (10)-(15) it can be written that

$$A(x) = R^2 \arccos\left(\frac{x_L - x}{R}\right) - (x_L - x)\sqrt{R^2 - (x_L - x)^2}, \quad (16)$$

which holds for $t_1 < t < t_2$.

The next step is analyzed the area $A$ for time interval $t_2 < t < t_3$, which is a little bit different. The illustration for this interval is given in Figure 5. In this interval the area $A$ can be found to be

$$A(x) = \pi R^2 - R^2 \arccos\left(\frac{x - x_L}{R}\right) + (x - x_L)\sqrt{R^2 - (x - x_L)^2}, \quad (17)$$

with similar way for Equation (16), which holds for $t_2 < t < t_3$.

For the last category when CL is leaving MFA, equations similar to Equation (16) and (17) can be simply derived,

which are given without detail proofs, with left side of MFA is defined as $x_R$,

$$A(x) = \pi R^2 - R^2 \arccos\left(\frac{x_R - x}{R}\right)$$
$$+ (x_R - x)\sqrt{R^2 - (x_R - x)^2}, \qquad (18)$$

which holds for $t_4 < t < t_5$, and

$$A(x) = R^2 \arccos\left(\frac{x - x_R}{R}\right)$$
$$- (x - x_R)\sqrt{R^2 - (x - x_R)^2}, \qquad (19)$$

which holds for $t_5 < t < t_6$.

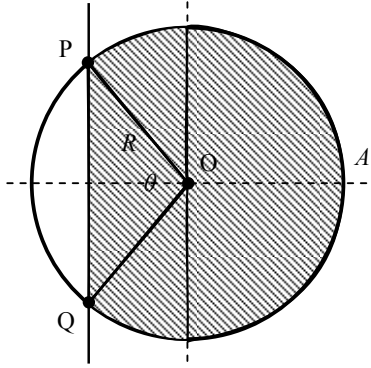

Figure 5. Area $A$ (shaded region) of CL as entering MFA for $t_2 < t < t_3$.

Equation (16)-(19) and results from the first category can summerized as follow

$$A(x) = \begin{cases} 0, & t < t_1 \\ R^2 \arccos\left(\frac{x_L - x}{R}\right) \\ -(x_L - x)\sqrt{R^2 - (x_L - x)^2}, & t_1 < t < t_2 \\ \pi R^2 - R^2 \arccos\left(\frac{x - x_L}{R}\right) \\ +(x - x_L)\sqrt{R^2 - (x - x_L)^2}, & t_2 < t < t_3 \\ 0, & t_3 < t < t_4 \quad (20) \\ \pi R^2 - R^2 \arccos\left(\frac{x_R - x}{R}\right) \\ +(x_R - x)\sqrt{R^2 - (x_R - x)^2}, & t_4 < t < t_5 \\ R^2 \arccos\left(\frac{x - x_R}{R}\right) \\ -(x - x_R)\sqrt{R^2 - (x - x_R)^2}, & t_5 < t < t_6 \\ 0, & t > t_6 \end{cases}$$

which holds for time $t$.

Finally the time $t_1$ until $t_6$ must be defined in order to have a valid time intervals. Since the LC moves with constant velocity $v$ and width of MFA is $w$, then it can be found that

$$t_2 = t_1 + \frac{D_l}{2v}$$
$$t_3 = t_1 + \frac{D_l}{v}$$
$$t_4 = t_1 + \frac{w}{v} \qquad , \qquad (21)$$
$$t_5 = t_1 + \frac{w}{v} + \frac{D_l}{2v}$$
$$t_6 = t_1 + \frac{w}{v} + \frac{D_l}{v}$$

with $t_1$ is the starting time of appeareance of induced EMF.

### *Induced EMF profile*

Induced EMF profile can found by applying Equation (20) into Equation (4), which is simply derivation of area $A$ with respect to time $t$. Using derivative of arrcos [6] it will be obtained that

$$\frac{d}{dt}\left[\arccos\left(\pm\frac{x_i - x}{R}\right)\right] = \pm\frac{v}{\left[R^2 - (x_i - x)^2\right]^{\frac{3}{2}}} \qquad (22)$$

and

$$\frac{d}{dt}\left[\pm(x_i - x)\sqrt{R^2 - (x_i - x)^2}\right] =$$
$$\mp v\sqrt{R^2 - (x_i - x)^2}\left[1 + \frac{(x_i - x)^2}{R^2 - (x_i - x)^2}\right], \qquad (23)$$

with $i = L, R$. Implementing Equation (22) and (23) for Equation (20) will give following functions. These functions are written separately only for simplicity.

$$\varepsilon(t < t_1) = 0, \qquad (24.a)$$

$$\varepsilon(t_1 < t < t_2) = \frac{BR^2v}{\left[R^2 - (x_L - x)^2\right]^{\frac{3}{2}}}$$
$$- vB\sqrt{R^2 - (x_L - x)^2}\left[1 + \frac{(x_L - x)^2}{R^2 - (x_L - x)^2}\right], \qquad (24.b)$$

$$\varepsilon(t_2 < t < t_3) = \frac{BR^2v}{\left[R^2 - (x - x_L)^2\right]^{\frac{3}{2}}}$$
$$- Bv\sqrt{R^2 - (x - x_L)^2}\left[1 + \frac{(x - x_L)^2}{R^2 - (x - x_L)^2}\right], \qquad (24.c)$$

$$\varepsilon(t_3 < t < t_4) = 0, \qquad (24.d)$$

$$\varepsilon(t_4 < t < t_5) = -\frac{BR^2v}{\left[R^2 - (x_R - x)^2\right]^{\frac{3}{2}}}$$
$$+ Bv\sqrt{R^2 - (x_R - x)^2}\left[1 + \frac{(x_R - x)^2}{R^2 - (x_R - x)^2}\right], \qquad (24.e)$$

$$\varepsilon(t_5 < t < t_6) = -\frac{BR^2 v}{\left[R^2 - (x - x_R)^2\right]^{\frac{3}{2}}}$$
$$+ Bv\sqrt{R^2 - (x - x_R)^2}\left[1 + \frac{(x - x_R)^2}{R^2 - (x - x_R)^2}\right], \quad (24.f)$$

$$\varepsilon(t > t_6) = 0, \quad (24.g)$$

with subsitution of

$$x(t) = (t - t_1)v \quad (25)$$

to make the equations as fully function of time $t$.

### Results and discussion

Plot of Equation (24.a)-(24.g) with Equation (25) is given in Figure 6 and the comparation from experiment from others [3] is given in Figure 7.

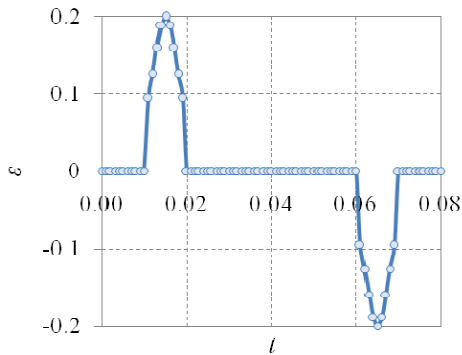

Figure 6. Plot of typical induced EMF from Equation (24.a)-(24.g) with Equation (25).

Following parameters are used for Figure 6: $w = 0.05$, $D_l = 0.1$, $v = 1$, $B = 10^{-3}$, $x_L = 0.005$, $x_R = 0.055$, $t_0 = 10^{-3}$, $t_0 = 0$, $t_1 = 0.01$, $t_2 = 0.015$, $t_3 = 0.02$, $t_4 = 0.06$, $t_5 = 0.065$, and $t_7 = 0.07$.

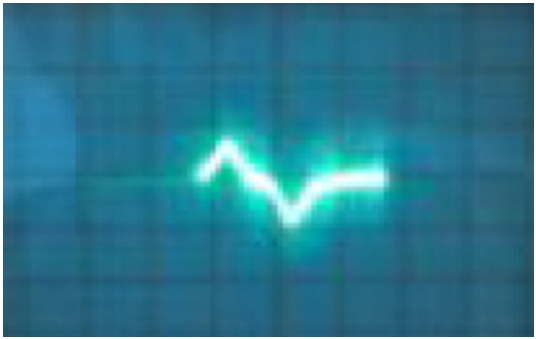

Figure 7. A typical results of reported experiment [3].

Comparation of Figure 6 and 7 can assure us that our induced EMF profile similar to the observed results reported by others [3].

Profile in Figure 6 depends on parameters such as $w$, $v$, $D_l$, and $B$ as it can be seen from Equation (20) and (24.a)-(24.g) with Equation (25).

### Conclusion

Derivation of induced EMF profile of a circular loop that is passing a limited rectangular area with perpendicular uniform magnetic field has been performed. The result from the derivation is similar to the reported observation.